\documentclass{ws-ijmpa}
\usepackage[super,compress]{cite}
\usepackage{graphicx}

\newcommand{\MET}{\ensuremath{E_\text{T}^\text{miss}}}

\begin{document}

\markboth{Y.Fang, M.Kumar, B.Mellado, Y.Zhang, M.Zhu }{Impact of additional bosons on the exploration of the Higgs boson at the LHC}

%
\catchline{}{}{}{}{}
%

\title{Impact of additional bosons on the exploration of the Higgs boson at the LHC}

\author{Yaquan Fang}
\address{Institute of High Energy Physics, Chinese Academy of Sciences, Beijing, China.}
\author{Mukesh Kumar}
\address{School of Physics, University of the Witwatersrand, Johannesburg, Wits 2050, South Africa. \\
National Institute for Theoretical Physics, School of Physics and Mandelstam Institute for
Theoretical Physics, University of the Witwatersrand, Johannesburg, Wits 2050, South Africa.
}
\author{Bruce Mellado}
\address{School of Physics, University of the Witwatersrand, Johannesburg, Wits 2050, South Africa.}
\author{Yu Zhang}
\author{Maosen Zhu}
\address{Institute of High Energy Physics, Chinese Academy of Sciences, Beijing, China.\\
University of Chinese Academy of Sciences, Beijing, China.}

\maketitle

\begin{history}
\received{Day Month Year}
\revised{Day Month Year}
\end{history}

\begin{abstract}
The LHC is making strides in the exploration of the properties of the newly discoverd Higgs boson, $h$. In Refs.~\cite{vonBuddenbrock:2015ema,Kumar:2016vut,vonBuddenbrock:2016rmr} the compatibility of the proton-proton data reported in the Run I period with the presence of a heavy scalar, $H$, with a mass around 270\,GeV and its implications were explored. This boson would decay predominantly to $H\rightarrow Sh$, where $S$, is a lighter scalar boson. The production cross-section of $pp\rightarrow H(\rightarrow Sh) + X$ is considerable and it would significantly affect the inclusive rate of $h$. The contamination from this new production mechanism would depend strongly on the final state used to measure the rate of $h$. The contamination in the rate measurement of $Vh(\rightarrow b\overline{b}), V=Z,W$ is estimated to be small. This statement does not depend strongly on assumptions made on the decay of $S$. 

\keywords{Higgs boson, Scalar bosons, BSM physics}
\end{abstract}

\ccode{PACS numbers:}


\section{Introduction}	
\label{intro}

With the observation of a Higgs-like scalar, $h$,~\cite{Englert:1964et,Higgs:1964ia,Higgs:1964pj,Guralnik:1964eu,ATLAS:2012gk,CMS:2012gu} at the Large Hadron Collider (LHC) the field of particle physics has dived into the study of its properties. The physics program of the LHC with the increase of the data set to 3-4 ab$^{-1}$ of integrated luminosity, will provide a unique opportunity to study the strength and structure of the couplings of the Higgs boson to particles in the Standard Model, SM, and, possibly, to particles beyond. 

In Refs.~\cite{vonBuddenbrock:2015ema,Kumar:2016vut,vonBuddenbrock:2016rmr} the compatibility of the results reported by the LHC experiments using proton-proton Run I data with a scalar boson, $H$, of a mass around 270\,GeV  and the corresponding phenomenology has been discussed. While the amount of results reported with Run II data remains limited, first studies are available that shed light of the compatibility of this hypothesis with new data has been reported at Ref.~\cite{vonBuddenbrock:2017bqf}. This heavy boson would decay predominantly into $h$ and another scalar boson, $S$. In Refs.~\cite{Kumar:2016vut,vonBuddenbrock:2016rmr} $S$ is considered a Higgs-like scalar with an additional interaction to Dark Matter, DM. 

The production cross-section of $pp\rightarrow H\rightarrow Sh$ would be at the level of 20\,pb at 13\,TeV. This represents a significant production mechanism of $h$,\footnote{The total cross-section of the SM Higgs boson in proton-proton collisions at 13\,TeV center of mass is 55\,pb.} where the Higgs boson would be accompanied by the decays of the $S$ boson. Assuming that the $S$ boson is Higgs-like, the Higgs boson would be expected to be produced with anomalously large jet and charged lepton multiplicity.  As a result, the contamination from this production mechanism depends strongly on the final state used for the isolation of the Higgs boson signal. It is important to note that the compatibility of the data so far with $H$ does not depend on the assumptions made on the nature of $S$ so far. 

The contamination from the $pp\rightarrow H\rightarrow Sh$ process may be significant when measuring the Higgs boson rate inclusively using the $h\rightarrow\gamma\gamma, ZZ(\rightarrow 4\ell)$ decays. This potential contamination could be strongly reduced with the implementation of  certain topological requirements. Two measurements of the Higgs boson rate are particularly important here: $h\rightarrow W^{(*)}W^{(*)}\rightarrow \ell\nu\ell\nu$ and $Vh(\rightarrow b\overline{b}), V=Z,W$. The former is isolated with the application of jet vetoes and the latter requires large transverse momenta of the weak boson while requiring limited jet activity to suppress top-backgrounds. It is predicted that the Higgs boson rates measured with these two final states be suppressed relative to the prediction from the SM compared to the inclusive rates measured using the $h\rightarrow\gamma\gamma, ZZ(\rightarrow 4\ell)$ decays. 

In this note a quantitative statement is made with regards to the potential contamination of the  $pp\rightarrow H\rightarrow Sh$ process to the measurement performed in the search for $Vh(\rightarrow b\overline{b}), V=Z,W$. The contamination on the measurement of the rate of  the Vector Boson Fusion (VBF) mechanism and $V(\rightarrow jj)h$ with the diphoton decay is also evaluated. The model dependence of these statements is discussed.

\section{The phenomenological framework}
\label{sec:pheno}

We consider an effective Lagrangian approach with the introduction of two hypothetical real scalars, 
$H$ and $\chi$, which are beyond the SM (BSM) in terms of its particle spectrum, as discussed in Refs.~\cite{vonBuddenbrock:2015ema,vonBuddenbrock:2016rmr}. This effective model can also be used to study other phenomenology associated with the Higgs boson physics.  
The formalism considers heavy scalar boson production though gluon-gluon fusion ($gg$F), which then decays into the SM Higgs and a pair of $\chi$ particles. As before, $\chi$ is considered as a DM candidate and therefore a source of missing transverse energy (\MET).  The transverse momentum of $h$ will be distinct from that of the direct production via $gg$F.

The required vertices for these studies are:
\begin{align}
\mathcal{L}_{H} =& -\frac{1}{4}~\beta_{g} \kappa_{_{hgg}}^{\text{SM}}~G_{\mu\nu}G^{\mu\nu}H
+\beta_{_V}\kappa_{_{hVV}}^{\text{SM}}~V_{\mu}V^{\mu}H,  \label{vh} \\
\mathcal{L}_{\text{Y}} =& -\frac{1}{\sqrt{2}}~\Big[y_{_{ttH}}\bar{t} t H + y_{_{bbH}} \bar{b} b H\Big],\\ 
\mathcal{L}_{\text{T}} =& -\frac{1}{2}~v\Big[\lambda_{_{Hhh}}Hhh + \lambda_{_{h\chi\chi}}h\chi\chi + \lambda_{_{H\chi\chi}}H\chi\chi\Big], \\
\mathcal{L}_{\text{Q}} =& -\frac{1}{2}\lambda_{_{Hh\chi\chi}}Hh\chi\chi - \frac{1}{4} \lambda_{_{HHhh}}HHhh 
-\frac{1}{4}\lambda_{_{hh\chi\chi}}hh \chi\chi \notag \\
& - \frac{1}{4} \lambda_{_{HH\chi\chi}}HH\chi\chi, \label{vq}
\end{align} 
where $\beta_g = y_{ttH}/y_{tth}$ is the scale factor with respect to the SM Yukawa top coupling for $H$, and it is therefore
used to tune the effective $gg$F coupling. A similar factor $\beta_V$ is used for $VVH$ couplings. 
The complete set of these new interactions are added to the SM Lagrangian, ${\cal{L}_{\text{SM}}}$, and thus the final Lagrangian is 
$\mathcal{L} = \mathcal{L}_{\text{SM}} + \mathcal{L}_{\text{BSM}}$,
where $\mathcal{L}_{\text{BSM}}$ contains the terms beyond the SM interactions which is given by 
\begin{align}
\mathcal{L}_{\text{BSM}} =&\, \frac{1}{2} \partial_\mu \chi \partial^\mu \chi - \frac{1}{2} m_\chi^2 \chi \chi
 + \frac{1}{2} \partial_\mu H \partial^\mu H \notag \\
 & - \frac{1}{2} m_H^2 H H 
 + \mathcal{L}_{H} + \mathcal{L}_\text{Y} + \mathcal{L}_\text{T} + \mathcal{L}_\text{Q}.\label{lag}
\end{align}

Here, we should note that $\chi$ only interacts with the SM Higgs boson and the postulated heavy scalar $H$ -- not with the SM fermions and gauge bosons. We also require that $\chi$ is stable by imposing the appropriate symmetry conditions. Since we assume $\chi$ to be a DM candidate,      
there are non-negligible constraints on the associated parameters of the vertices that come from the relic density of DM and the DM-nuclei 
inelastic scattering cross sections. In addition to this, constraints arise from limits on the invisible BR of the SM Higgs boson. These leave a narrow choice of the mass of the DM candidate, $m_\chi \sim m_h/2$, as well as the parameter $\lambda_{h\chi\chi} \sim [0.0006-0.006]$. We also assume that $m_H$ would lie in the range,  $2 m_h < m_H < 2 m_t$ to forbid the $H\to t t$ decay, as well as keep the $H\to h\chi\chi$ decay on-shell. 

In an effective field theory approach, we do not consider the actual origin of the $Hh\chi\chi$ coupling. One can assume 
that this effective interaction is mediated by the scalar particle $S$ which will then decay in the mode $S\to\chi\chi$. This inclusion of $S$ can 
open up various new possibilities in terms of search channels and phenomenology, as discussed in Refs.~\cite{vonBuddenbrock:2015ema,Kumar:2016vut,vonBuddenbrock:2016rmr}. The vertices defined above
(in Eqs.~\ref{vh} to \ref{vq}) will be modified appropriately with $S$ as an intermediate scalar and not as a DM candidate. $S$ is a scalar particle with various decay modes, therefore having all possible decays to other SM particles. As a result, the symmetry requirements for a gauge invariant set of vertices in the 
Lagrangian is different. With the mass range $m_h \lesssim m_S  \lesssim m_H - m_h$ and $m_S > 2m_\chi$, new possibilities for the processes in these studies include $pp \to H\to Sh$ as well as $pp \to H \to hh$, considering the available
spectrum of $m_S$ and the associated coupling parameters. There is a possibility to introduce a $HSS$ vertex in the study, which participates further in a $H \to S S$  decay channel (similar to $H \to h h$). 

Following the effective theory approach, and after Electro-Weak symmetry breaking, the Lagrangian for singlet real scalar 
$S$ can be written as:
\begin{equation}
{\cal L}_{S} = {\cal L}_\text{K} + {\cal L}_{SVV^\prime}
 + {\cal L}_{Sf\bar f} + {\cal L}_{hHS} + {\cal L}_{S\chi},
\end{equation}
where
\begin{equation}
{\cal L}_\text{K} =\, \frac{1}{2} \partial_\mu S \partial^\mu S - \frac{1}{2} m_S^2 S S, \label{lsk}
\end{equation}
\begin{align}
{\cal L}_{SVV^\prime} =&\, \frac{1}{4} \kappa_{_{Sgg}} \frac{\alpha_{s}}{12 \pi v} S G^{a\mu\nu}G_{\mu\nu}^a
+ \frac{1}{4} \kappa_{_{S\gamma\gamma}} \frac{\alpha}{\pi v} S F^{\mu\nu}F_{\mu\nu} \notag \\
& + \frac{1}{4} \kappa_{_{SZZ}} \frac{\alpha}{\pi v} S Z^{\mu\nu}Z_{\mu\nu} 
  + \frac{1}{4} \kappa_{_{SZ\gamma}} \frac{\alpha}{\pi v} S Z^{\mu\nu}F_{\mu\nu} \notag \\
& + \frac{1}{4} \kappa_{_{SWW}} \frac{2 \alpha}{\pi s_w^2 v} S W^{+\mu\nu}W^{-}_{\mu\nu}, \label{lsvv}
\end{align}
\begin{equation}
{\cal L}_{Sf\bar f} =\, - \sum_f \kappa_{_{Sf}} \frac{m_f}{v} S \bar f f, \label{lsf}
\end{equation}
\begin{align}
{\cal L}_{HhS} =&\, -\frac{1}{2}~v\Big[\lambda_{_{hhS}} hhS + \lambda_{_{hSS}} hSS + 
\lambda_{_{HHS}} HHS \notag \\ & + \lambda_{_{HSS}} HSS + \lambda_{_{HhS}} HhS\Big], \label{lsh}
\end{align}
\begin{equation}
{\cal L}_{S\chi} =\, -\frac{1}{2}~v~\lambda_{_{S\chi\chi}} S\chi\chi -\frac{1}{2} \lambda_{_{SS\chi\chi}} SS\chi\chi. \label{lsc} 
\end{equation}

Here $V, V^\prime \equiv g, \gamma, Z \,\text{or}\, W^\pm$ and $W^\pm_{\mu\nu} = D_\mu W^\pm_\nu - D_\nu W^\pm_\mu $,
$D_\mu W^\pm_\nu = \left[ \partial_\mu \pm i e A_\mu \right] W^\pm_\nu$. Other possible self interaction terms for $S$
are neglected here since they are not of any phenomenological interest for our studies. Hence the total effective Lagrangian is:
\begin{equation}
\cal{L}_{\text{tot}} = \cal{L}_\text{SM} + \cal{L}_\text{S}. \label{lagt}
\end{equation} 

It is interesting to note that the choice of narrow mass range for $S$, $m_S\in [m_h, m_H-m_h]$ 
provides an opportunity to see various phenomenological aspects of the model in contrast to $h$. The mass range for $S$ may help to understand rates for a Higgs-like scalar in different possible production or decay modes too. An important feature to keep in mind is that all decay modes of $S$ (i.e. $S$ into jets, vector bosons, leptons, DM etc.) are possible.  This features are explored to estimate the contamination of the process $pp \to H\to Sh$ in the study of the Higgs boson couplings. 

\section{Tools and event selections}
\label{sec:tools_evsel}

Samples were generated, showered and hadronized using \texttt{Pythia 8}~\cite{Sjostrand:2007gs}. The detector level simulation is performed with reasonably chosen parameters 
using \texttt{Delphes}~\cite{deFavereau:2013fsa} and jets were clustered using \texttt{FastJet}~\cite{Cacciari:2011ma} 
with the anti-$k_T$ algorithm~\cite{Cacciari:2008gp} using the distance parameter, $R=0.4$.
The factorization and normalization scales for the signal samples are fixed to the $H$ mass. The cross-section for  $pp\rightarrow H(\rightarrow Sh) + X$ is set to 10\,pb for 13\,TeV proton-proton center of mass, which is a conservative benchmark. The cross-sections for the SM production mechanisms come from the Higgs cross-section working group. 

Here a number of final states are considered as important benchmarks for the study of Higgs boson couplings at the LHC. The contamination with respect to the relevant SM production mechanisms are evaluated for the following final states:
\begin{itemize}
\item [{\bf 1.}] The production via VBF with $h\rightarrow\gamma\gamma$.
\item [{\bf 2.}] The associated production $Vh, V=Z,W$ with $V\rightarrow jj$ and $h\rightarrow\gamma\gamma$. 
\item [{\bf 3.}] The associated production $Vh$ with $h\rightarrow b\overline{b}$.
\end{itemize}

A number of requirements are imposed on the phase-space. When two photons are present in the final state the following requirements are implemented~\cite{Aad:2014eha,Aad:2014lwa}:
\begin{equation}
p_T^{\gamma}>25\text{\,GeV}, \left|\eta^{\gamma}\right|<2.37,
\end{equation}
where $p_T^{\gamma}$ and $\eta^{\gamma}$ are the photon transverse momenta and pseudorapidity, respectively. It is required that the region of $1.37<\left|\eta^{\gamma}\right|<1.52$ be excluded. It is also required that:
\begin{equation}
\frac{ p_T^{\gamma 1(2)}}{m_{\gamma\gamma} } > 0.35(0.25), 
\end{equation}
where  $p_T^{\gamma 1}$ and $p_T^{\gamma 2}$ are the leading and subleading photon transverse momenta, respectively, and $m_{\gamma\gamma}$ is the di-photon invariant mass. The diphoton invariant mass, $m_{\gamma\gamma}$, is required to be in the range of $120<m_{\gamma\gamma}<130$\,GeV.

Hadronic jets are required to have a transverse momentum greater than 25\,GeV and be in the range $\left|\eta\right|<4.5$.

\subsection{VBF, $h\rightarrow\gamma\gamma$}
\label{sec:vbf}
In addition to the aforementioned requirements on the photons, at least two hadronic jets are required such that~\cite{Aad:2014eha}:
\begin{equation}
\label{eq:vbf}
m_{jj}>400\text{\,GeV}, \Delta\eta_{jj}>2.8, \Delta\Phi_{\gamma\gamma, jj}>2.8\text{\,rad},
\end{equation}
where $\Delta\Phi_{\gamma\gamma, jj}$ is the azimuthal angle difference between the diphoton and dijet systems.

\subsection{ $Vh, V\rightarrow jj, h\rightarrow\gamma\gamma$}
\label{sec:vhyy}
In addition to the aforementioned requirements on the photons, at least two hadronic jets are required such that:
\begin{equation}
\label{eq:vh}
60<m_{jj}<110\text{\,GeV}, \Delta\eta_{\gamma\gamma,jj}<1, p_{Tt\gamma\gamma}>70\text{\,GeV}, \Delta\eta_{jj}<3.5, 
\end{equation}
where $m_{jj}$ is the leading dijet invariant mass, $p_{Tt\gamma\gamma}$ is the component of the diphoton transverse momentum that is orthogonal to the axis defined by the difference between the two photon momenta~\cite{Ackerstaff:1997rc,Vesterinen:2008hx}, and $\Delta\eta_{jj}$ is the rapidity difference between the leading jets.

\subsection{$Vh$, $h\rightarrow b\overline{b}$}
\label{sec:vhbb}
The search for the associated production $Vh$ with $h\rightarrow b\overline{b}$ is typically performed according to the number of charged leptons (electrons or muons) in the final state: 0-lep, 1-lep and 2-lep categories. A number of requirements are imposed such that the transverse momentum of the weak boson be large while ensuring reduced jet activity. The event selection used here follows that reported in Ref.~\cite{ATLAS-CONF-2016-091}. It is important to note that multivariate techniques are used in addition to the event selections described below. The impact of the multivariate discriminants is not covered here.
\subsubsection{0-lep}
\label{sec:0lep}
A lepton veto is applied consisting of no electrons with $p_{T}>7$ GeV, $|\eta|<2.47$  and no muons with $p_{T}>7$ GeV, $|\eta|<2.7$. The event is required to have less or equal than three jets, where the jet is defined as $p_{T}>20$ GeV in $|\eta|<2.5$ or $p_{T}>30$ GeV in $2.5<|\eta|<4.5$. Two b-tag jets are required with  $p_{T}>20$ GeV and $|\eta|<2.5$, where the leading b-tag jet is required to have $p_{T} > 45$\,GeV. The scalar sum of all jets is denoted as $S_T$, where $S_T>120, 150$\,GeV for events with exactly two or three jets, respectively. It is required that $\MET>150$\,GeV. The following topological conditions are imposed
\begin{equation}
\label{eq:0lep}
\Delta^{min}\Phi_{\MET,jet}>20^{\circ}, \Delta\Phi_{\MET,h}>120^{\circ}, \Delta\Phi_{b1, b2}<90^{\circ},
\end{equation}
where $\Delta^{min}\Phi_{\MET,jet}$ is the minimum azimuthal angle distance between $\MET$ and all jets. $\Delta\Phi_{\MET,h}$ and $\Delta\Phi_{b1, b2}$ are the azimuthal angle difference between the $\MET$ and the Higgs boson candidate, and the two b-tag jets, respectively. 
\subsubsection{1-lep}
\label{sec:1lep}
Exactly one charged lepton is required with with $p_{T}>25$ GeV ,$\left|\eta\right|<$2.47 for electron or $|\eta|<2.7$ for muons. A veto on additional charged leptons with $7<p_{T}<25$\,GeV is applied. Events with an electron are required to have $\MET>30$\,GeV. Less or equal than three jets are required, where jets are defined with two b-tags as in Sec.~\ref{sec:0lep}. The transverse momentum of the system made by the lepton and the $\MET$ is required to be $p_{TV}>150$\,GeV.

\subsubsection{2-lep}
\label{sec:2lep}

Exactly two charged leptons that are same flavor and opposite sign with $p_{T} > 7$\,GeV are required such that the dilepton invariant mass lie in the range $71<m_{ll}<121$. The event is required to have exactly two b-tag jets, as defined in Sec.~\ref{sec:0lep}. Two categories are made, depending on the weak boson candidate transverse momentum: $p_{TV}<150$\,GeV and $p_{TV}>150$\,GeV.

\section{Results}
\label{sec:results}

Here the basic kinematic distributions are presented together with the expected contamination from the $H\rightarrow Sh$ decays into the corners of the phase-space defined in Sec.~\ref{sec:tools_evsel}.

\begin{figure}[t]
\centerline{\includegraphics[width=6.5cm]{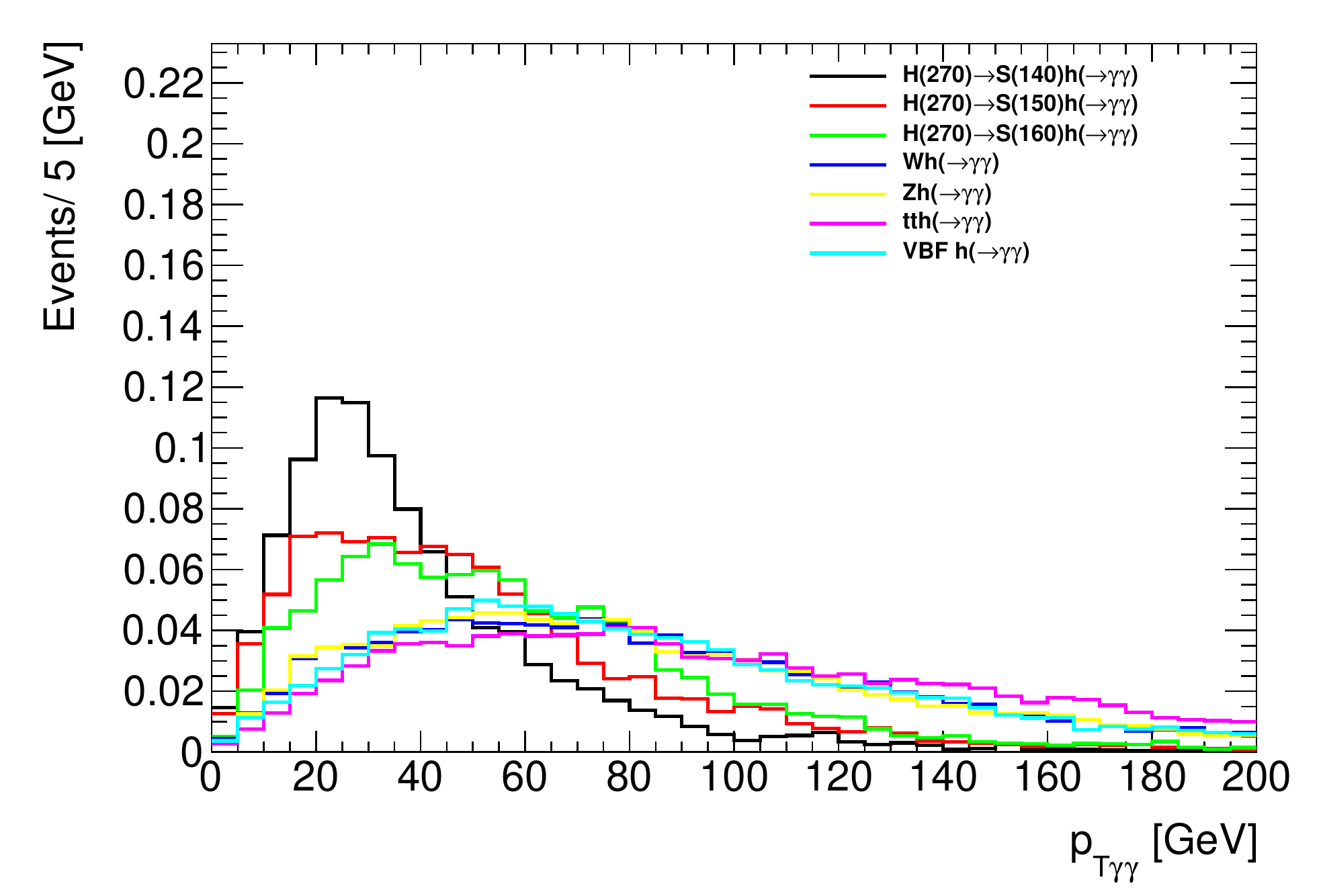}
\includegraphics[width=6.5cm]{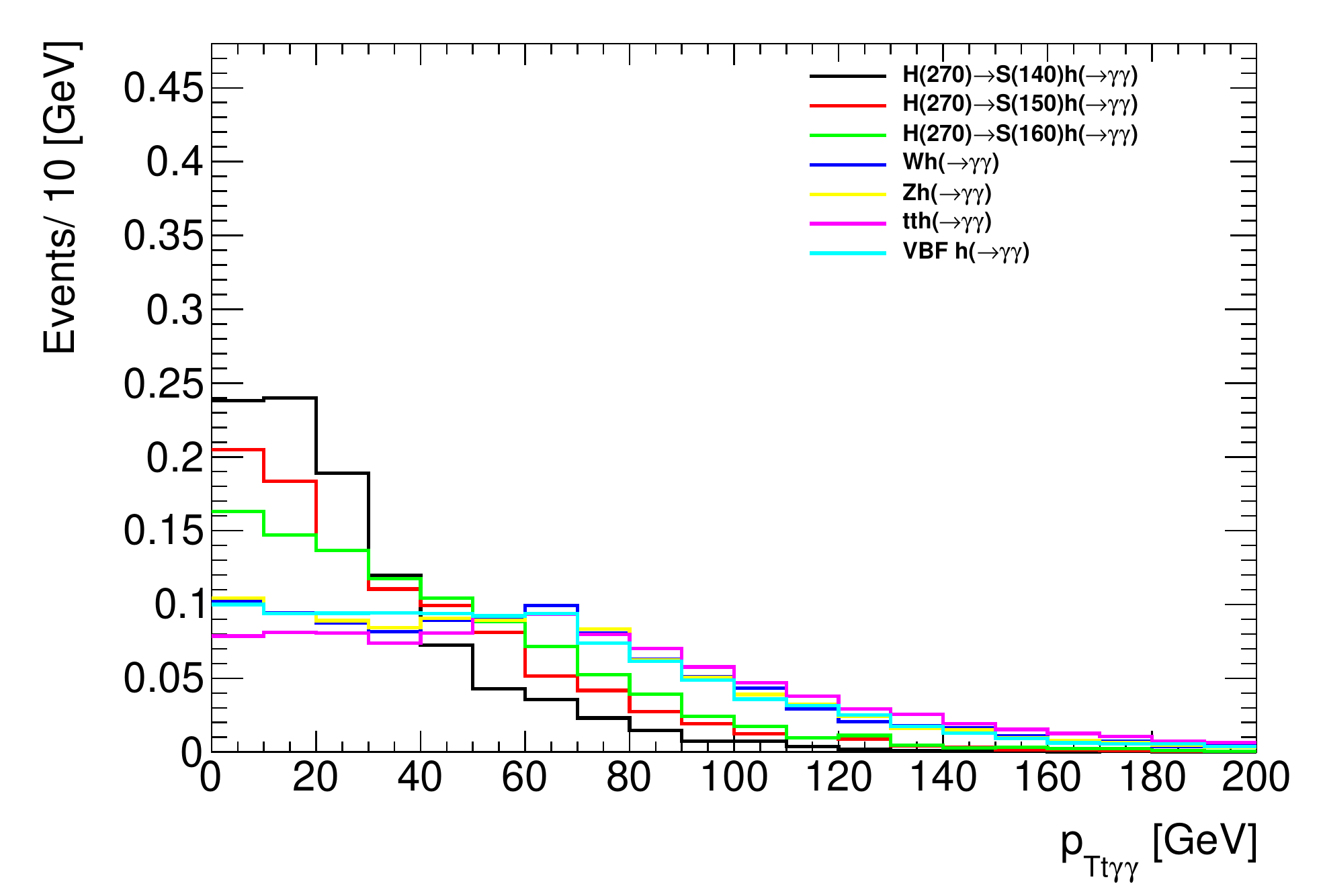}}
\caption{Distributions of the diphoton transverse momentum (left) and $p_{Tt}$ (right). See text. \label{fig:ptyy}}
\end{figure}

Figure~\ref{fig:ptyy} displays the diphoton transverse momentum and $p_{Tt}$ (see Sec.~\ref{sec:vhyy}). The distributions are normalized to unity. The kinematics of the production of $H\rightarrow Sh$ with $h\rightarrow\gamma\gamma$ for representative masses of $H$, $m_H=270$\,GeV and $S=140, 150, 160$\,GeV are compared with that of the relevant SM processes. For the choice of masses made here, the diphoton system tends to be softer than for the associated SM processes considered. 

\begin{figure}[t]
\centerline{
\includegraphics[width=6.5cm]{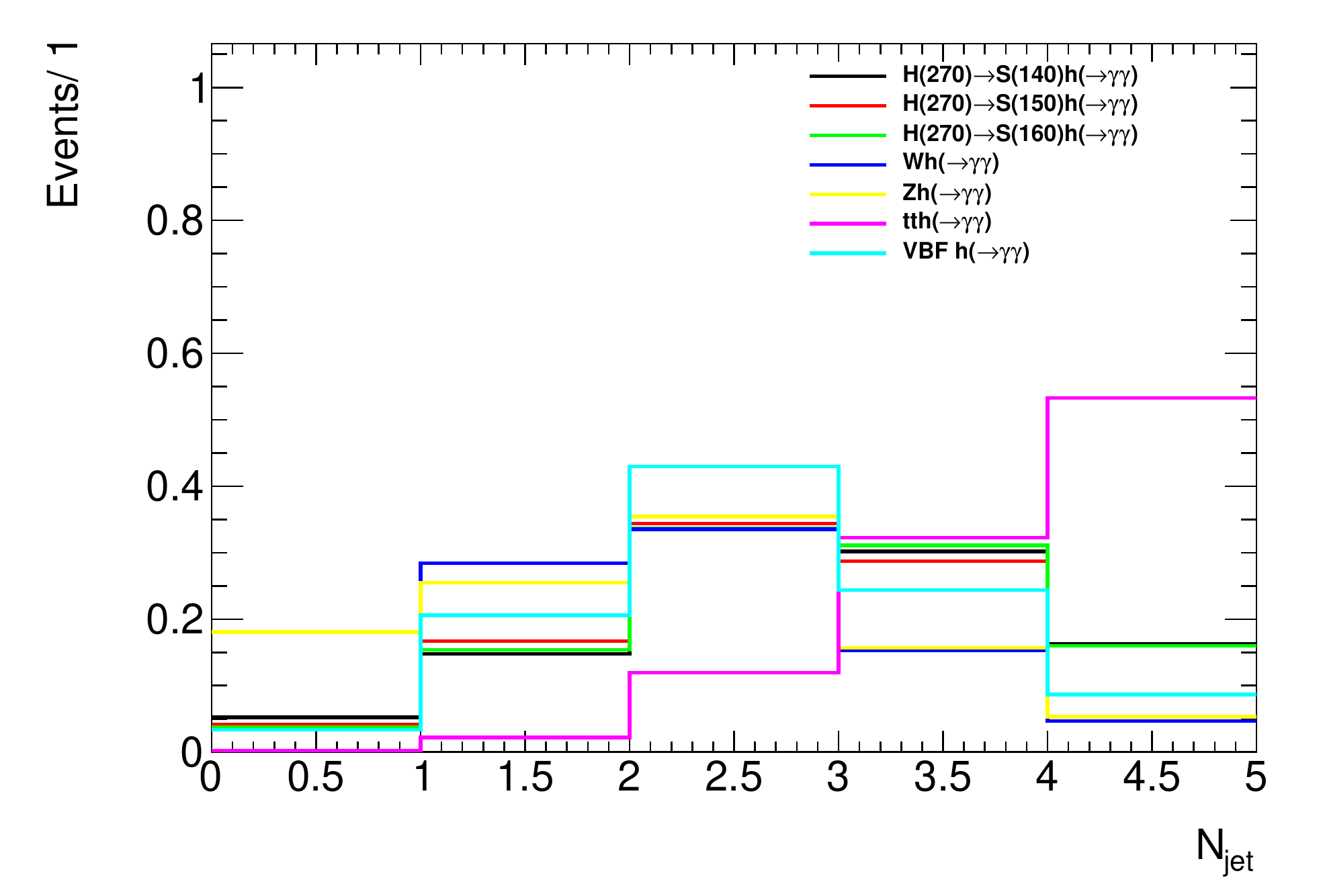}
\includegraphics[width=6.5cm]{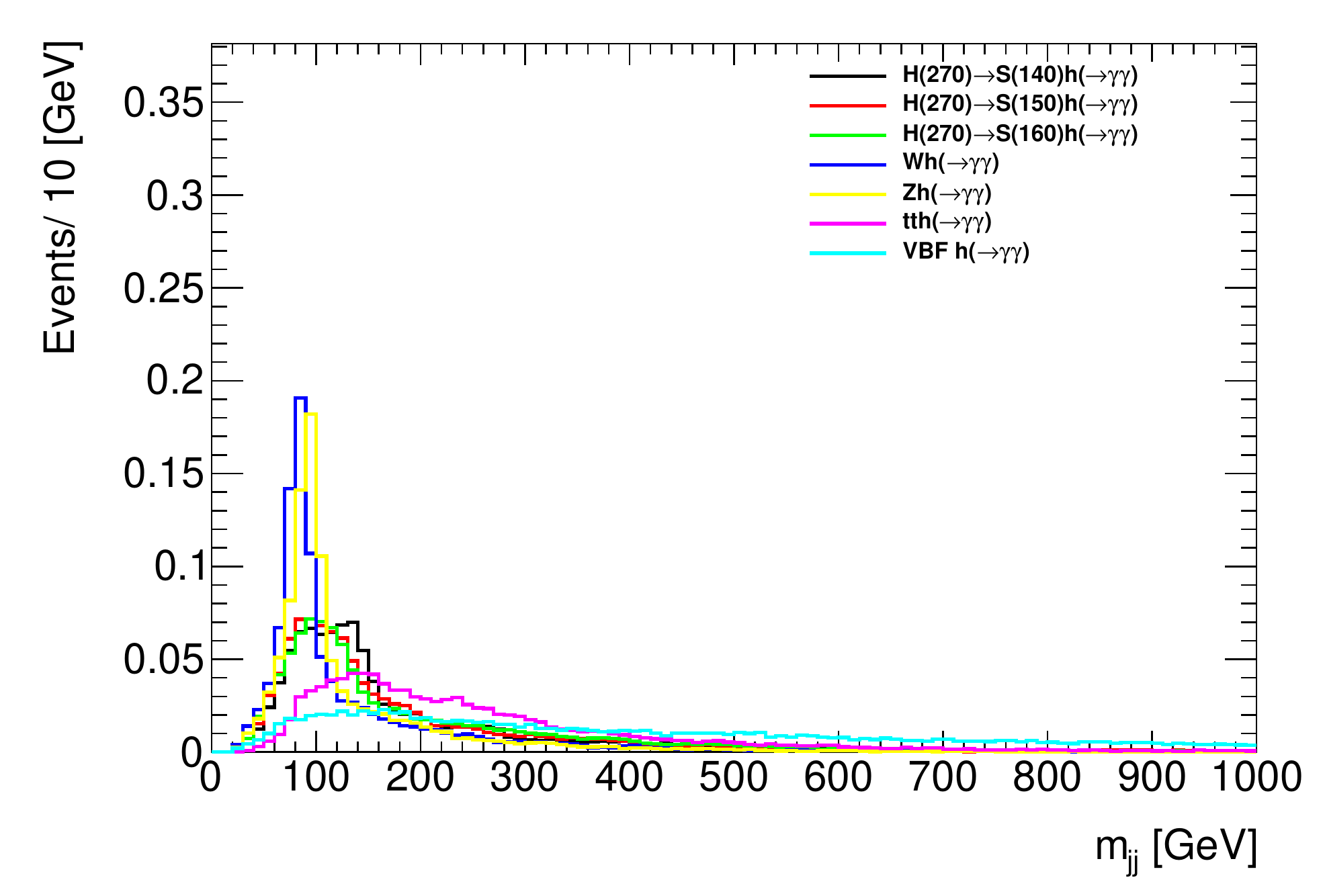}}
\centerline{
\includegraphics[width=6.5cm]{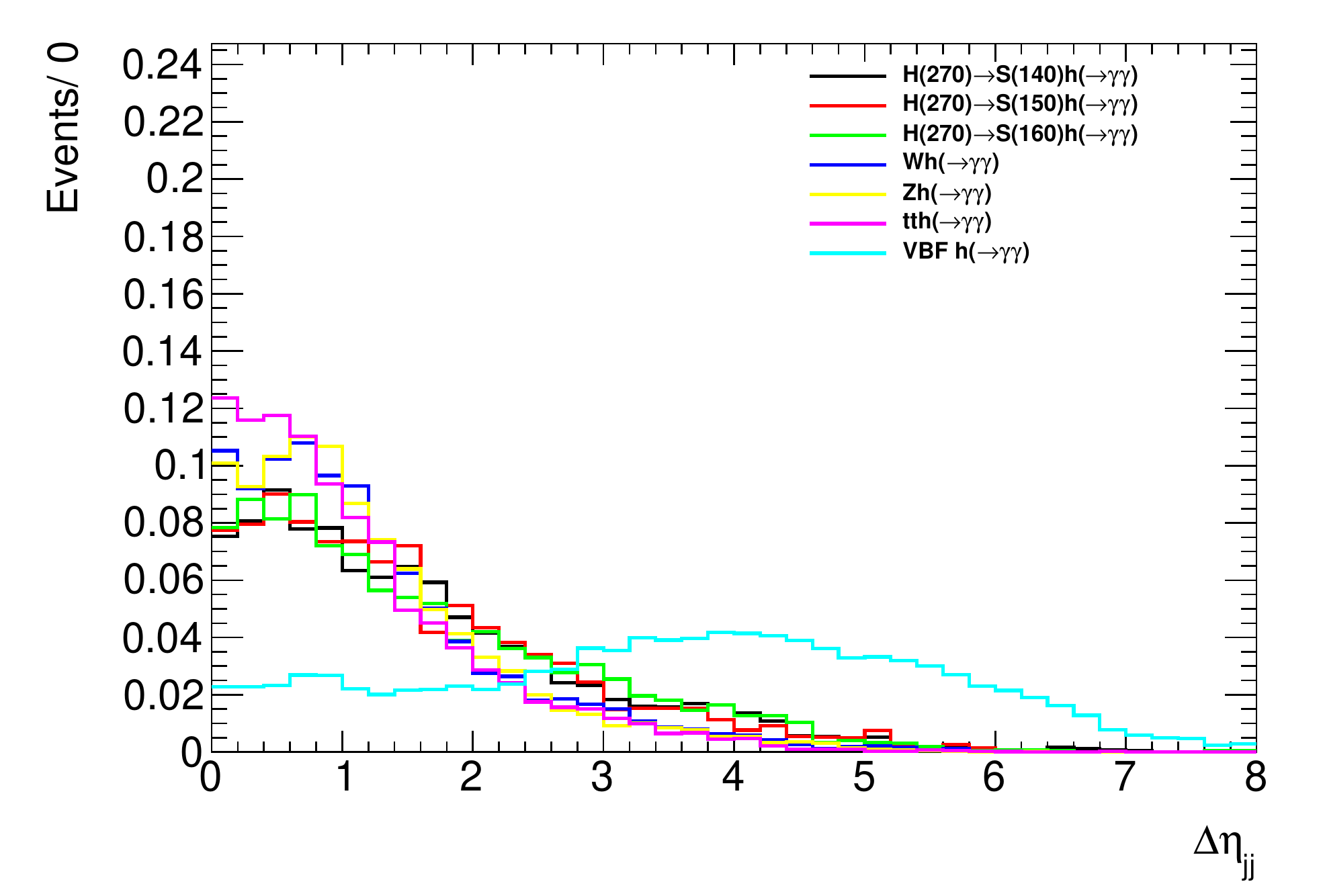}
\includegraphics[width=6.5cm]{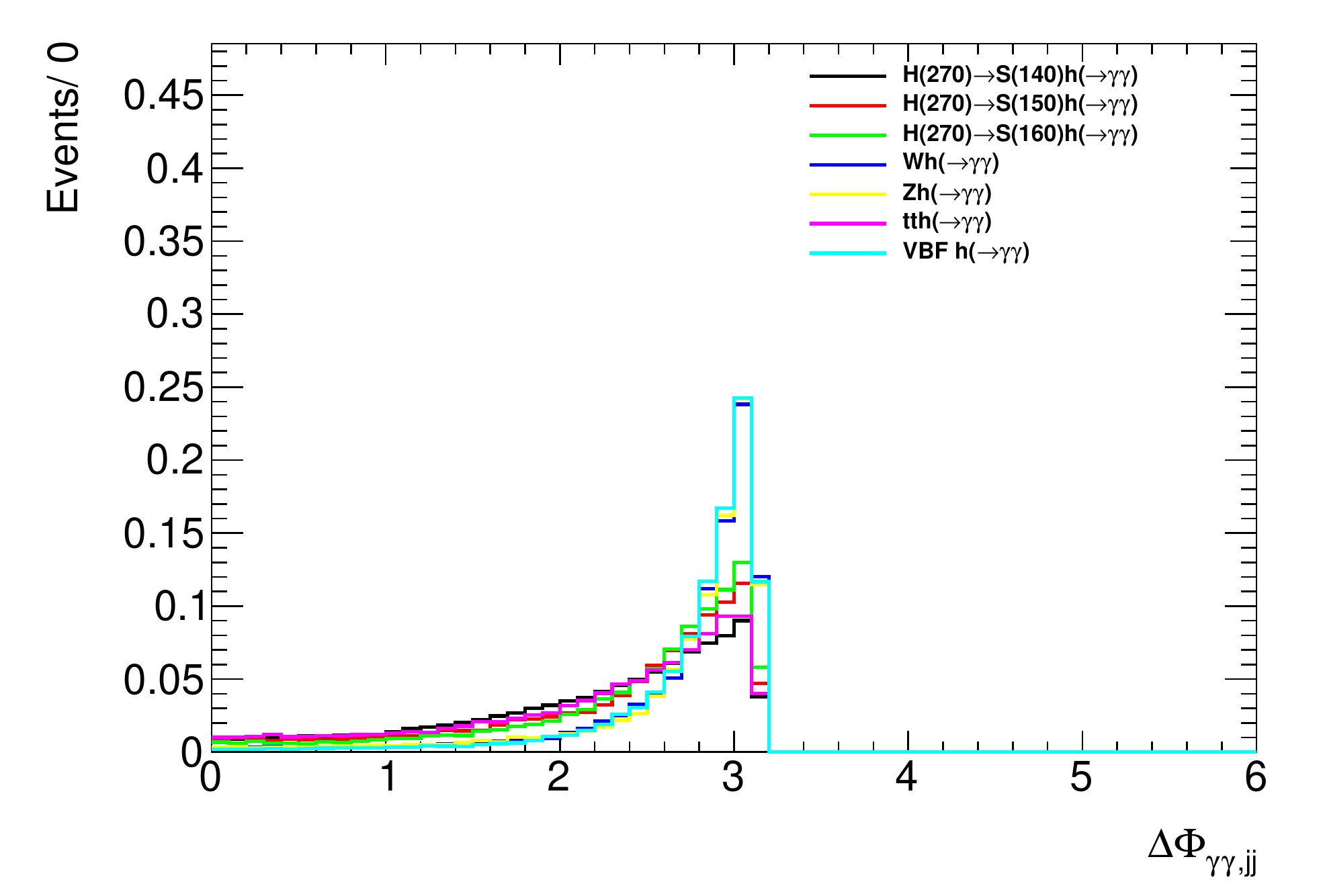}
}
\caption{Kinematic distributions related to jets usind the Higgs boson diphoton decay. Shown are the jet multiplicity (upper left), the dijet invariant mass (upper right), the dijet rapidity difference (lower left) and the azimuthal difference between the diphoton and dijet systems (lower right).  \label{fig:jetsyy}}
\end{figure}

Figure~\ref{fig:jetsyy} displays jet related kinematics. The $H\rightarrow Sh$ production mechanism tends to have small jet veto survival probability, whereas a significant fraction of the events have exactly one jet. The jet multiplicity differential distribution has a strong model dependence that is not studied here. Similar discussion applies to the rest of the observables displayed in Fig.~\ref{fig:jetsyy}. The dependence on the mass of $S$ is not strong.

\begin{table}[ph]
\tbl{Expected yields for 36\,fb$^{-1}$ of integrated luminosity for 13\,TeV proton-proton center of mass energy for the VBF, $Vh$ event selections described in Secs.~\ref{sec:vbf} and~\ref{sec:vhyy}. The $H\rightarrow Sh$ production mechanism is compared to SM associated production mechanisms. Errors correspond to the statistical error of the MC sample.}
{\begin{tabular}{@{}ccc@{}} \toprule
Production mechanism          & VBF $h\rightarrow\gamma\gamma$      &$Vh, V\rightarrow jj, h\rightarrow\gamma\gamma$\\
\hline
$H(270)\rightarrow S(140)h(\rightarrow\gamma\gamma)$          &2.86$\pm$0.07          &0.16$\pm$0.02          \\
\hline
$H(270)\rightarrow S(150)h(\rightarrow\gamma\gamma)$          &1.94$\pm$0.06          &1.14$\pm$0.04          \\
\hline
$H(270)\rightarrow S(160)h(\rightarrow\gamma\gamma)$          &2.89$\pm$0.07          &1.97$\pm$0.06          \\
\hline
$Wh(\rightarrow\gamma\gamma)$               &0.22$\pm$0.01          &1.90$\pm$0.03          \\
\hline
$Zh(\rightarrow\gamma\gamma)$               &0.14$\pm$0.01          &1.31$\pm$0.02          \\
\hline
$tth(\rightarrow\gamma\gamma)$              &0.09$\pm$0.00          &0.22$\pm$0.01		\\
\hline
VBF $h(\rightarrow\gamma\gamma)$              &25.81$\pm$0.20         &0.30$\pm$0.02          \\
\hline
\end{tabular} \label{tab:vbfvh}}
\end{table}

Table~\ref{tab:vbfvh} shows the yields for 36\,fb$^{-1}$ of integrated luminosity for 13\,TeV proton-proton center of mass energy for the $H\rightarrow Sh$ production mechanism and SM associated production mechanisms. The yields correspond to the event selections for VBF, $Vh$ with $h\rightarrow\gamma\gamma$ described in Sec.~\ref{sec:tools_evsel}. Relevant here is the relative yield of the $H\rightarrow Sh$ production mechanism compared to those of the SM. 

The distinctive topological features of the VBF mechanism (see Fig.~\ref{fig:jetsyy}) allows this mechanism to remain dominant with respect to other production mechanisms considered here. Assuming that the cross-section of $pp\rightarrow H(\rightarrow Sh) + X$ is set to 10\,pb, the contamination is about 11\%. Needless to say, this estimate depends on the assumption made here that $S$ decays as a Higgs-like boson. The corresponding contamination from $H\rightarrow Sh$ in the hadronic $Vh$ search depends significantly on the $S$ mass. The contamination grows from a  small contribution with $m_S=140$\,GeV to about 53\% of the SM contribution when $m_S$ approaches twice the $V$ masses. The experimental precision to date is not sufficient to establish contamination of the size discussed here. For instance, the most recent results from the CMS experiment (see Ref.~\cite{CMS:2017rli}) using the diphoton decay report a signal strength of $\mu_{VBF}=0.6^{+0.6}_{-0.5}$ and $\mu_{Vh,had}=4.1^{+2.5}_{-2.3}$. Data collected beyond Run II will be needed to determine if significant contamination is present in these final states. 

\begin{figure}[t]
\centerline{
\includegraphics[width=6.5cm]{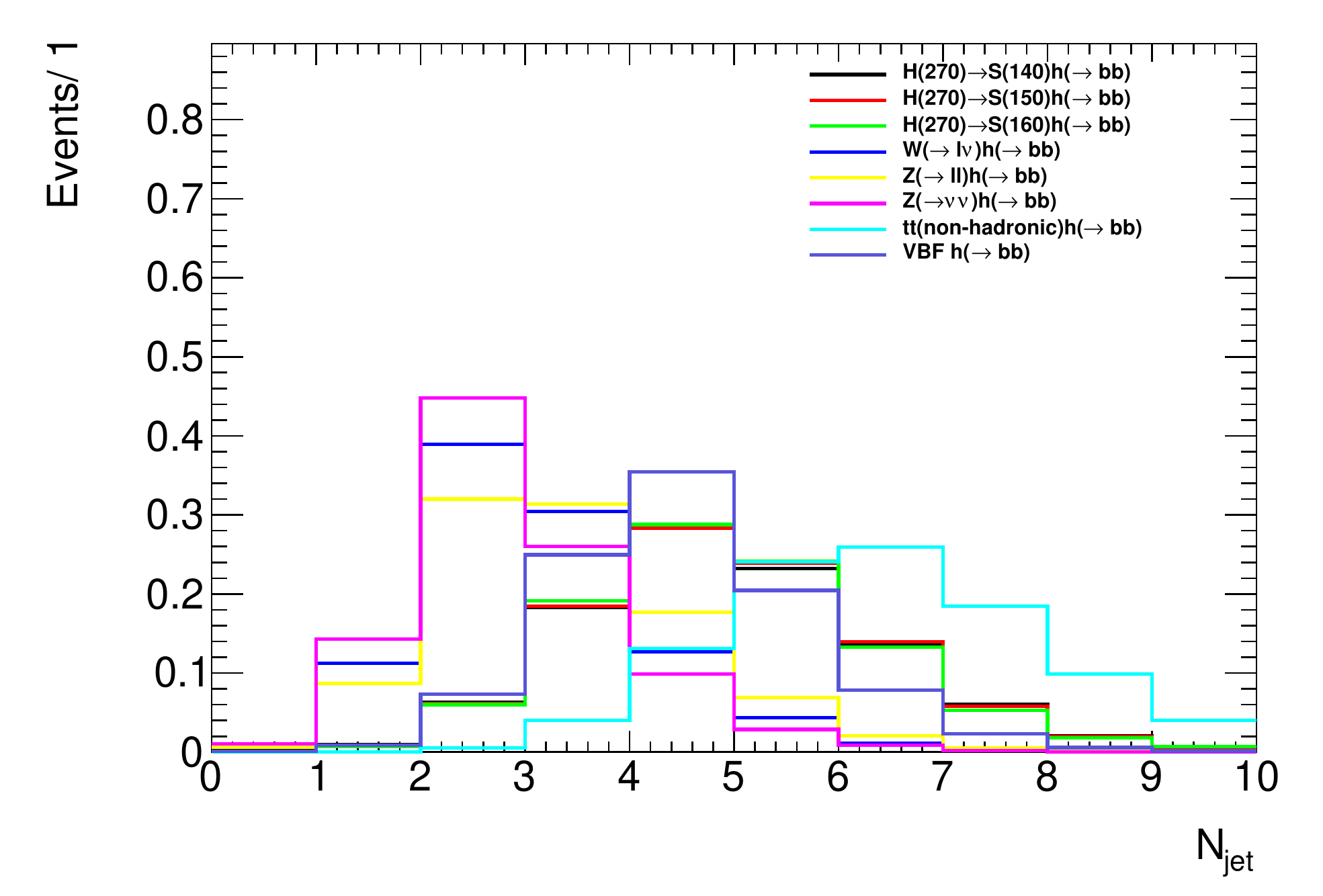}
\includegraphics[width=6.5cm]{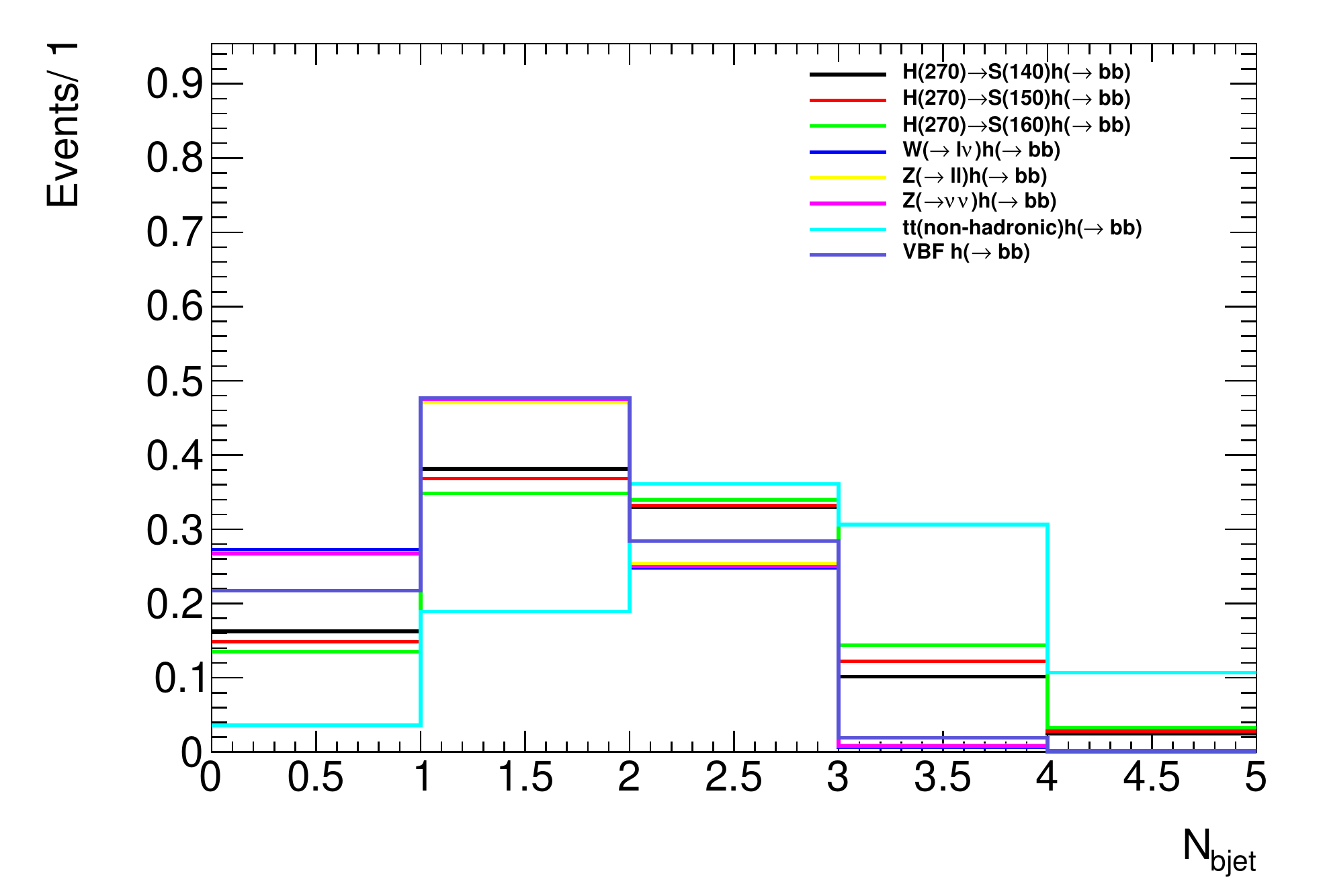}}
\caption{Jet (left) and b-jet (right) multiplicities using the $h\rightarrow b\overline{b}$ decay.  \label{fig:njetbb}}
\end{figure}

\begin{figure}[t]
\centerline{
\includegraphics[width=6.5cm]{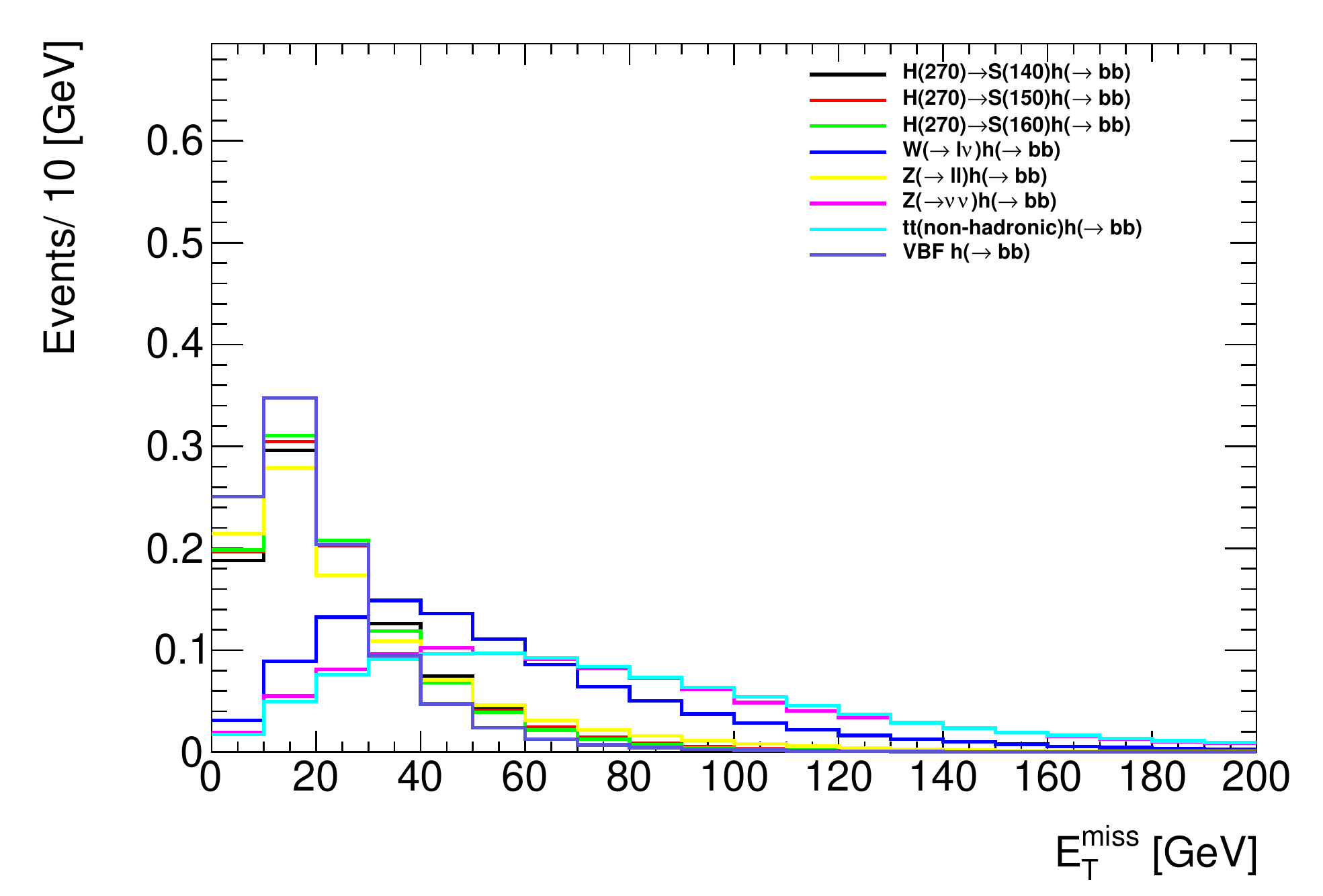}
\includegraphics[width=6.5cm]{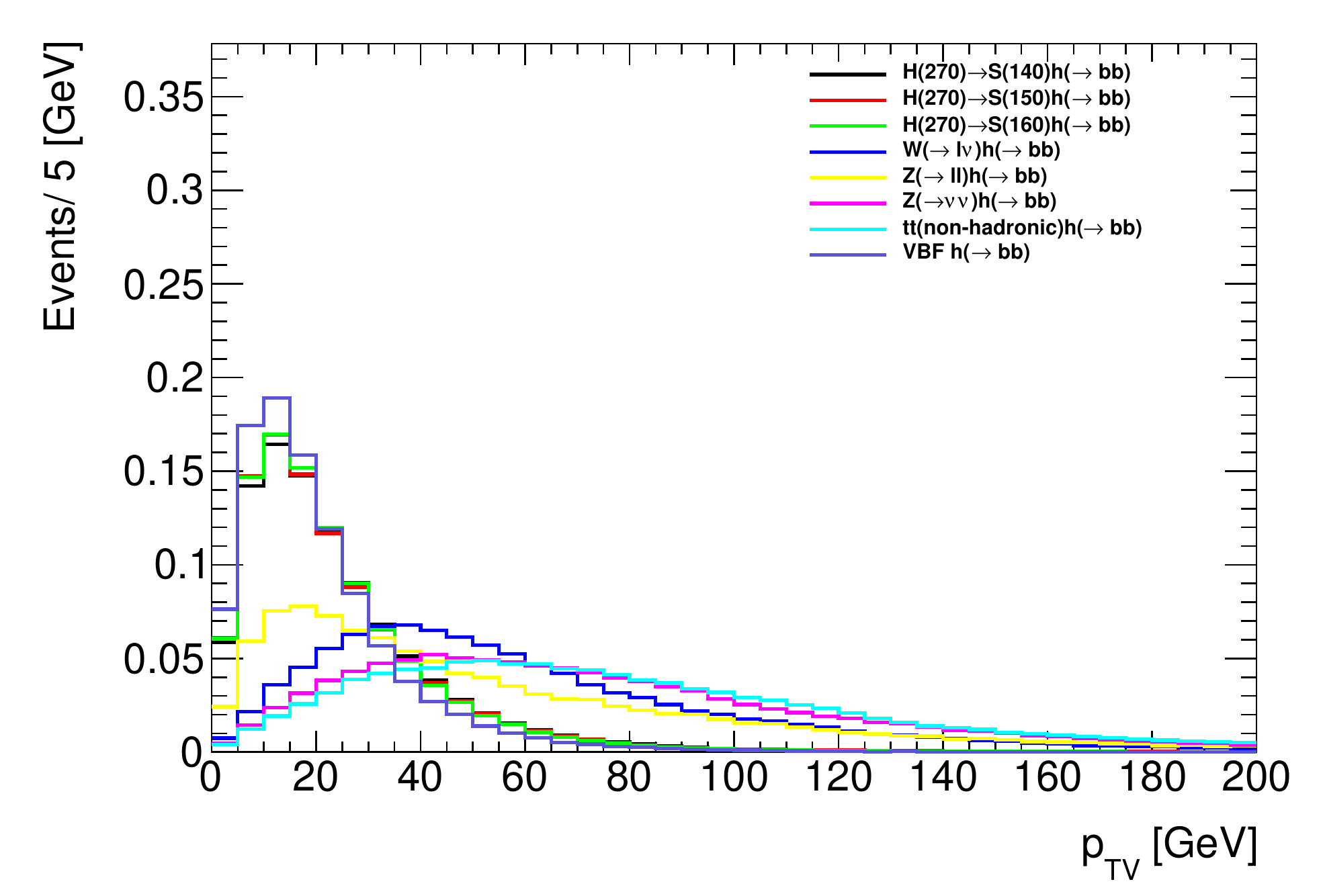}
}
\caption{Transverse momentum of the $V$ candidate in the search for $Vh, h\rightarrow b\overline{b}$ (see text).  \label{fig:ptbb}}
\end{figure}

Figure~\ref{fig:njetbb} displays the jet and b-jet multiplicities using the production mechanisms discussed above with the $h\rightarrow b\overline{b}$ decay. The $Vh$ production mechanism shows little jet activity. This is in contrast to the Higgs boson production in association with top quarks. The $\rightarrow Sh$ production mechanism displays more jet activity compared to the $Vh$ production mechanism.

Figure~\ref{fig:ptbb} shows the transverse momentum of the $V$ candidate in the event selections described in Sec~\ref{sec:vhbb}. The plot on the left corresponds to the $\MET$ distribution corresponding to the 0-lep category. The plot on the right corresponds to the $p{TV}$ distribution in 1-lep and 2-lep events. The distributions are shown after the requirements of physical objects described in Sec.~\ref{sec:vhbb} and before topological requirements are imposed. The distinctive feature of the associated production $Vh$ in the SM is the large transverse momentum of the $V, h$ candidates in the absence of large jet multiplicity (see the event selections in Sec.~\ref{sec:vhbb}). This is in contrast with the kinematics of the $H\rightarrow Sh$ production mechanism. The transverse momentum of the decay products in $H\rightarrow Sh$ are bound by the masses of the bosons. The corresponding spectrum of the $\MET$ in this production mechanism is significantly softer compared to what is expected in the SM. 

\begin{table}[ph]
\tbl{Expected yields for 36\,fb$^{-1}$ of integrated luminosity for 13\,TeV proton-proton center of mass energy for the $Vh, h\rightarrow b\overline{b}$ event selections described in Sec.~\ref{sec:vhbb}. The $H\rightarrow Sh$ production mechanism is compared to SM associated production mechanisms. Errors correspond to the statistical error of the MC sample.}
{\begin{tabular}{@{}ccccc@{}} \toprule
Production mechanism           &0-lep      &1-lep      &2-lep, low $p_{TV}$        &2-lep, high $p_{TV}$\\
\hline
$H(270)\rightarrow S(140)h(\rightarrow bb)$           &$<$0.10                &0.21$\pm$0.15          &26.74$\pm$1.67         &$<$0.10                \\
\hline
$H(270)\rightarrow S(150)h(\rightarrow bb)$           &0.52$\pm$0.23          &2.39$\pm$0.50          &60.87$\pm$2.52         &$<$0.10                \\
\hline
$H(270)\rightarrow S(160)h(\rightarrow bb)$           &$<$0.10                &0.31$\pm$0.18          &13.63$\pm$1.19         &$<$0.10                \\
\hline
$W(\rightarrow l\nu)h(\rightarrow bb)$              &21.10$\pm$1.01         &100.76$\pm$2.20        &$<$0.05                &$<$0.05                \\
\hline
$Z(\rightarrow ll)h(\rightarrow bb)$              &0.35$\pm$0.06          &2.41$\pm$0.15          &57.21$\pm$0.72         &18.84$\pm$0.41	     \\
\hline
$Z(\rightarrow \nu\nu)h(\rightarrow bb)$              &62.40$\pm$1.07         &$<$0.02                &$<$0.02                &$<$0.02                \\
\hline
VBF $h(\rightarrow bb)$               &$<$0.79                &$<$0.79                &$<$0.79                &$<$0.79                \\
\hline
$tt(non-had.) h(\rightarrow bb)$        &0.11$\pm$0.02          &0.56$\pm$0.05          &0.50$\pm$0.05          &0.04$\pm$0.01\\
\hline
\end{tabular} \label{tab:vhbb}}
\end{table}

Table~\ref{tab:vhbb} shows the expected yields for 36\,fb$^{-1}$ of integrated luminosity for 13\,TeV proton-proton center of mass energy for the $Vh, h\rightarrow b\overline{b}$ event selections described in Sec.~\ref{sec:vhbb}. The event selections for the 0-lep and 1-lep categories contain the stringent requirement that $p_{TV}>150$\,GeV. This naturally suppresses the contribution from the $H\rightarrow Sh$ production mechanism, as indicated in Tab.~\ref{tab:vhbb}. 

The 2-lep category classifies events according to $p_{TV}$, allowing events with $p_{TV}<150$\,GeV in the search. In the low $p_{TV}$ category the contamination can be considerable, of order of the SM, while in the high $p_{TV}$ it is negligible. According to Ref.~\cite{ATLAS-CONF-2016-091}, the relative sensitivity of the 2-lep, low $p_{TV}$ carries about 10\% of the sensitivity to the signal strength. As noted in Sec.~\ref{sec:vhbb}, in addition to the event selection described there, multivariate techniques are applied to distinguish between the Higgs boson signal and SM non-resonant backgrounds. One of the input variables is $p_{TV}$. This will further suppress the contribution from the $H\rightarrow Sh$ in this corner of the phase-space. However, this effect is not quantified here. Assuming $\sigma(pp\rightarrow H(\rightarrow Sh) + X)=10$\,pb, the potential impact on the $Vh$ signal strength measurement would be less than 10\%.

\section{Summary}
\label{sec:summary}

In Refs.~\cite{vonBuddenbrock:2015ema,Kumar:2016vut,vonBuddenbrock:2016rmr} the compatibility of the proton-proton data reported in the Run I period with the presence of a heavy scalar with a mass around 270\,GeV and its implications were explored. This boson would decay predominantly to $H\rightarrow Sh$, where $S$, is a lighter scalar boson. The production cross-section of $pp\rightarrow H(\rightarrow Sh) + X$ is considerable and it would significantly affect the inclusive rate of $h$. Here the potential contamination of the $H\rightarrow Sh$ production mechanism in Higgs boson searches that impose topological requirements is evaluated. This is done assuming that the $S$ boson has Higgs boson-like decays. 

The contamination is evaluated in the search for the Higgs boson via VBF and $Vh, V\rightarrow jj$ using the Higgs boson decay to two photons. Assuming $\sigma(pp\rightarrow H(\rightarrow Sh) + X)=10$\,pb, the contamination would be about 11\% and up to 53\%, respectively. This contamination  is expected to be model dependent. This aspect is not covered here. The level of accuracy of the measurements of the signal strenghts to date does not allow to draw conclusions. Data collected beyond Run II will be needed to determine if significant contamination is present in these final states.

The contamination is also evaluated in the search for $Vh, h\rightarrow b\overline{b}$. The SM predicts that the transverse momentum of the weak boson be large associated with moderate jet multiplicity. This feature strongly suppresses the contamination provided $p_{TV}>150$\,GeV. This statement is not expected to depend on the assumptions made on the decays of $S$ and it is mostly determined by the size of the mass of $H$ and $S$ considered here. In the 2-lep and low $p_{TV}$ category, the contamination can be as large as the contribution from the SM. However, this category has little influence on the sensitivity to the signal strength. Assuming $\sigma(pp\rightarrow H(\rightarrow Sh) + X)=10$\,pb, the impact on the measured signal strength would be less than 10\%.

\section*{Acknowledgments}
Y.F. wants to thank the Chinese Academy of Sciences Center for Excellence in Particle Physics. The High Energy Physics group of the University of the Witwatersrand is grateful for the support from the Wits Research Office, the National Research Foundation, the National Institute of Theoretical Physics and the Department of Science and Technology through the SA-CERN consortium.



\begin{thebibliography}{0}    


\bibitem{Englert:1964et} 
  F.~Englert and R.~Brout,
  Phys.\ Rev.\ Lett.\  {\bf 13}, 321 (1964)

\bibitem{Higgs:1964pj} 
  P.~W.~Higgs,
  Phys.\ Rev.\ Lett.\  {\bf 13}, 508 (1964)

\bibitem{Higgs:1964ia} 
  P.~W.~Higgs,
  Phys.\ Lett.\  {\bf 12}, 132 (1964)

\bibitem{Guralnik:1964eu} 
  G.~S.~Guralnik, C.~R.~Hagen and T.~W.~B.~Kibble,
  Phys.\ Rev.\ Lett.\  {\bf 13}, 585 (1964)

\bibitem{ATLAS:2012gk}
ATLAS  Collaboration (G.~Aad {\em et~al.}) {\em Phys.Lett.} {\bf
 B716}, 1  (2012)

\bibitem{CMS:2012gu}
CMS  Collaboration (S.~Chatrchyan {\em et~al.}), {\em Phys.Lett.}
 {\bf B716}, 30  (2012)


\bibitem{vonBuddenbrock:2015ema} 
  S.~von Buddenbrock {\it et al.},
  arXiv:1506.00612 [hep-ph].
 

\bibitem{Kumar:2016vut}
  M.~Kumar {\it et al.},
  J.\ Phys.\ Conf.\ Ser.\  {\bf 802} (2017) no.1,  012007
 

\bibitem{vonBuddenbrock:2016rmr}
  S.~von Buddenbrock {\it et al.},
  Eur.\ Phys.\ J.\ C {\bf 76} (2016) no.10,  580
  
\bibitem{vonBuddenbrock:2017bqf}
  S.~von Buddenbrock,
  arXiv:1706.02477 [hep-ph].


  
\bibitem{Sjostrand:2007gs} 
  T.~Sjostrand, S.~Mrenna and P.~Z.~Skands,
  Comput.\ Phys.\ Commun.\  {\bf 178}, 852 (2008)
 
 
 \bibitem{Aad:2014eha} 
  G.~Aad {\it et al.} [ATLAS Collaboration],
  Phys.\ Rev.\ D {\bf 90}, no. 11, 112015 (2014)
  
  \bibitem{Aad:2014lwa}
  G.~Aad {\it et al.} [ATLAS Collaboration],
  JHEP {\bf 1409} (2014) 112

\bibitem{deFavereau:2013fsa} 
  J.~de Favereau {\it et al.} [DELPHES 3 Collaboration],
  JHEP {\bf 1402}, 057 (2014)

\bibitem{Cacciari:2011ma} 
  M.~Cacciari, G.~P.~Salam and G.~Soyez,
  Eur.\ Phys.\ J.\ C {\bf 72}, 1896 (2012)


\bibitem{Cacciari:2008gp} 
  M.~Cacciari, G.~P.~Salam and G.~Soyez,
  JHEP {\bf 0804}, 063 (2008)
 
\bibitem{Ackerstaff:1997rc} 
  K.~Ackerstaff {\it et al.} [OPAL Collaboration],
  Eur.\ Phys.\ J.\ C {\bf 4}, 47 (1998)
  
\bibitem{Vesterinen:2008hx} 
  M.~Vesterinen and T.~R.~Wyatt,
  Nucl.\ Instrum.\ Meth.\ A {\bf 602}, 432 (2009)  
  
\bibitem{ATLAS-CONF-2016-091} 
  ATLAS Collaboration, ``Search for the Standard Model Higgs boson produced in association with a vector boson and decaying into a $b\overline{b}$ pair in $pp$ collisions at 13\,TeV using the ATLAS detector'',
  ATLAS-CONF-2016-091
  
  
\bibitem{CMS:2017rli} 
  CMS Collaboration, ``Measurements of properties of the Higgs boson in the diphoton decay channel with the full 2016 data set'',
  CMS-PAS-HIG-16-040
  
 
\end{thebibliography}
\end{document}